\documentclass{WileyMSP-template_modif}

\begin{document}

\pagestyle{fancy}
\rhead{\includegraphics[width=2.5cm]{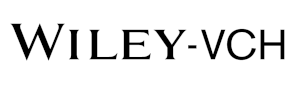}}

\title{Chemical evolution of the Universe\\
and its consequences for gravitational-wave astrophysics}
\maketitle

% Author: Please give full first and last names for authors and include * after the name of all corresponding authors

\author{Martyna Chru{\'s}li{\'n}ska*}
% \author{Author Two}
% \author{Author Three*}

% Dedication
\dedication{}

% Affiliations: Please provide adacemic titles (Prof. or Dr.) for all authors where applicable, and include an institutional email address for all corresponding authors
\begin{affiliations}
Dr. Martyna Chru{\'s}li{\'n}ska\\
Max Planck Institute for Astrophysics,  
Karl-Schwarzschild-Str. 1 
85741 Garching, Germany\\
% Address\\
Email Address:
mchruslinska@mpa-garching.mpg.de
% A. N. O. Author\\
% Address

\end{affiliations}

% Keywords: Please provide a minimum of three and a maximum of seven keywords, separated by commas

\keywords{gravitational waves, galaxies:  abundances, galaxies: star formation}

% Abstract should be written in the present tense and impersonal style (i.e., avoid we), and be at most 200 words long
\begin{abstract}

We are now routinely detecting gravitational waves (GW) emitted by merging black holes and neutron stars. Those are the afterlives of massive stars that formed all across the Universe - at different cosmic times and with different metallicities (abundances of elements heavier than helium).
\\
Birth metallicity plays an important role in the evolution of massive stars.
Consequently, the population properties of mergers are sensitive to the metallicity dependent cosmic star formation history (f$_{\rm SFR}$(Z,z)).
In particular, within the isolated formation scenarios (the focus of this paper), a strong low metallicity preference of the formation of mergers involving black holes was found. The origin of this dependence and its consequences are discussed. Most importantly, uncertainty in the f$_{\rm SFR}$(Z,z) (substantial even at low redshifts, especially at low metallicity) cannot be ignored in the models. This poses a significant challenge for the interpretation of the observed GW source population properties. Possible paths for improvements and the role of future GW detectors are considered.
\\
Recent efforts to determine f$_{\rm SFR}$(Z,z) and the factors that dominate its uncertainty are summarized. Many of those factors are related to the properties of galaxies that are faint and distant and therefore difficult to observe.
The fact that they leave imprint on the properties of mergers as a function of cosmic time makes future GW observations a promising (and complementary to electromagnetic observations) tool to study chemical evolution of galaxies.
\end{abstract}

\section{From massive stars to gravitational-wave sources}

Massive stars evolve into black holes (BH) and neutron stars (NS) within just a few to few tens of Myrs \cite{Hurley00, Woosley02}.
As such, BH/NS progenitors are only found in regions that are currently star forming and can be used as star formation rate tracers \cite{MadauDickinson14,EldridgeStanway22}.
In contrast, the compact objects resulting from their evolution gradually populate galaxies, reflecting their integrated (massive) star formation history.
  \begin{figure}[h!]
  \begin{minipage}[c]{0.7\textwidth}
    \includegraphics[width=1.\textwidth]{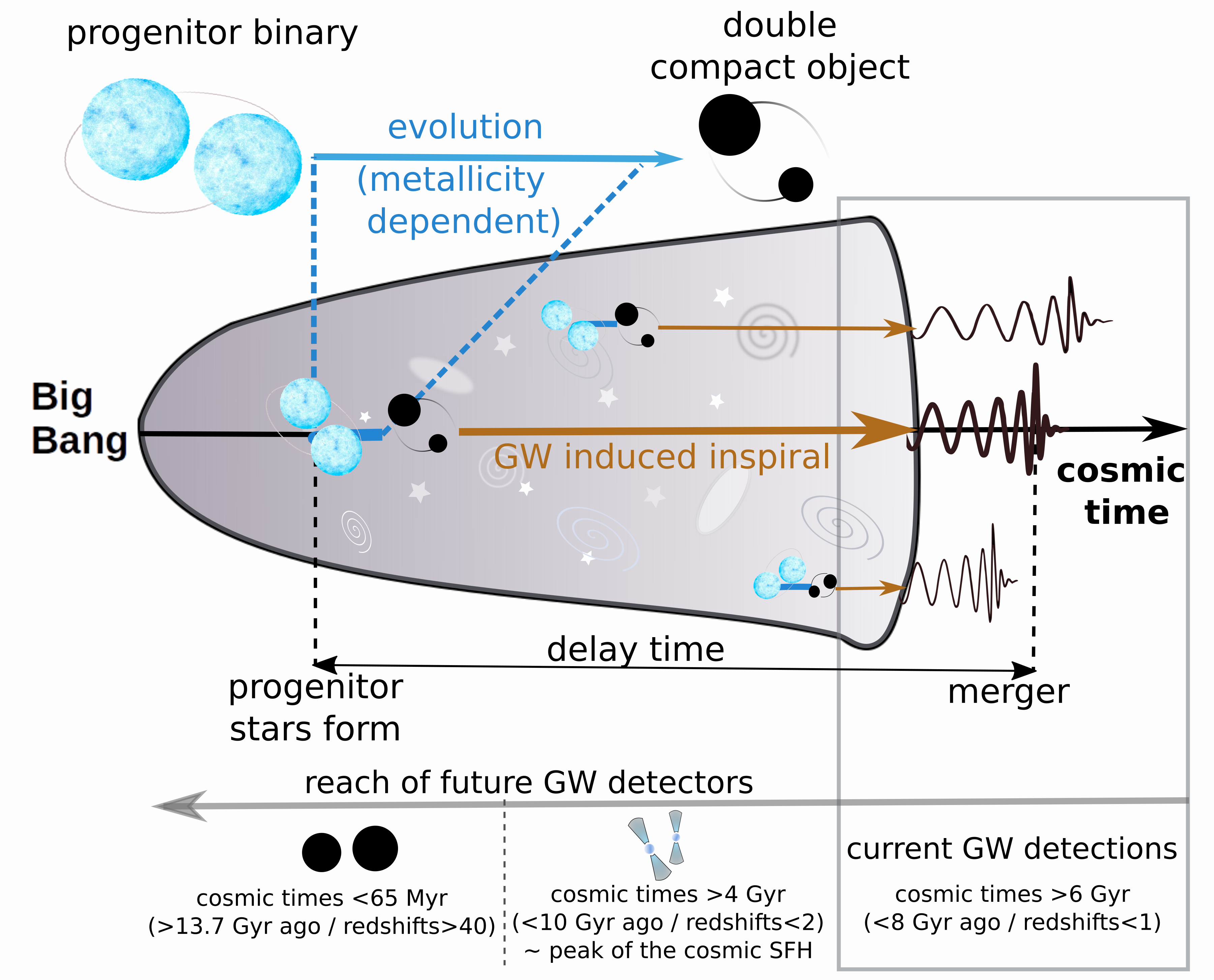}
  \end{minipage}\hfill
  \begin{minipage}[c]{0.3\textwidth}
    \caption{
Black hole/neutron star mergers observed with GW originate from massive stars that formed at different cosmic times, in different environments and with different metallicities. Even the locally observed population (probed with current detectors) may contain systems whose progenitors formed very early in the cosmic history.
The approximate range of cosmic times/redshifts spanned by current GW detections \cite{GWTC3} and the reach of the optimistic network of future GW detectors considered in \cite{BorhanianSathyaprakash22} (see Table 3 therein) for NS+NS and BH+BH mergers are indicated at the bottom.
    }             
    \label{fig: schematic GW population}
  \end{minipage}
\end{figure}
Even though massive stars rarely evolve without any companion stars(s) \cite{Sana12, Sana13,MoeDiStefano17}, only a small fraction of stellar BHs and NSs form binaries (either BH+BH, BH+NS or NS+NS systems, referred to as double compact objects - DCO further in this paper).
Mergers of such binaries are the main astrophysical source of gravitational waves (GW) that can be observed with current detectors. In fact, all of the astrophysical GW signals detected to date were interpreted as coming from DCO mergers \cite{GWTC2,GWTC3}.
DCO can form as a direct product of evolution of a stellar binary/higher-order multiple or when BHs and/or NSs encounter each other and form a bound system due to dynamical interactions.
The latter can happen at a non-negligible rate only in densely populated stellar regions -- such as nuclear/globular clusters.
Only DCOs that merge within a time shorter than the age of the Universe ($\tau_{0}\sim$14 Gyr) are potentially observable with ground-based detectors.
Regardless of the formation path, such binaries may form with a wide range of delay times (i.e.\ the time between the formation of the progenitor stars and the merger; \cite{MennekensVanbeveren16,Belczynski16,Lipunov17,Stevenson17,EldridgeStanwayTang19,vanSon21,Fishbach21}).
The important consequence of this fact is that the observed population of DCO mergers, even the local events, contains a mixture of systems formed throughout the entire cosmic history (see Figure \ref{fig: schematic GW population}).
As discussed in the next section, the contents of this mixture are set by both the evolution that leads to the formation of DCO mergers and the (metallicity-dependent) cosmic star formation history.

\subsection{Population modelling: the need for metallicity-dependent cosmic star formation history}

\begin{figure}[h]
%   \begin{minipage}[c]{0.6\textwidth}
    \includegraphics[width=1.\textwidth]{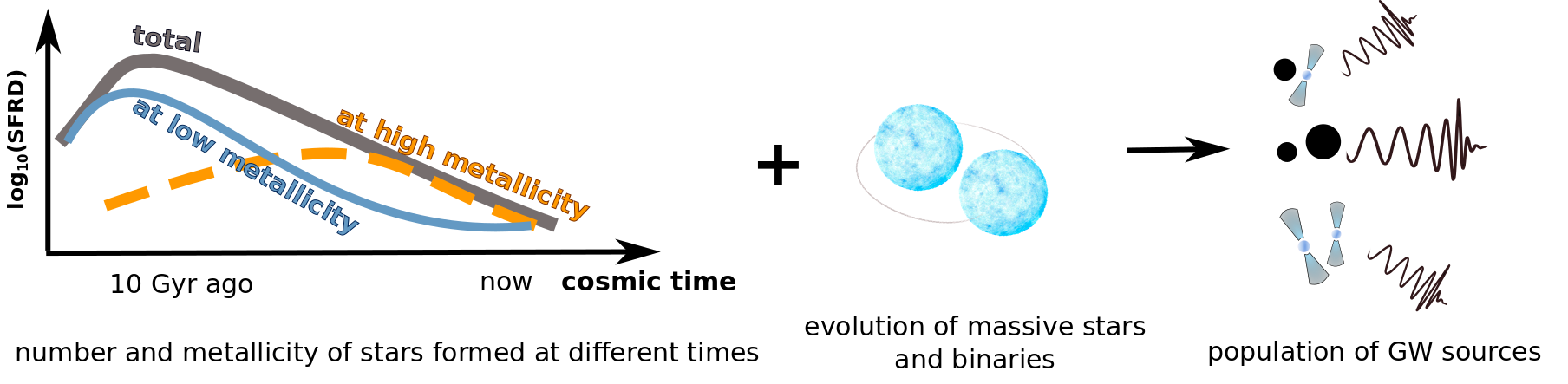}
%   \end{minipage}\hfill
%   \begin{minipage}[c]{0.4\textwidth}
    \caption{
Key parts of the modelling of the population of DCO mergers. Metallicity dependent cosmic star formation history (left)
needs to be combined with a model describing evolution of massive stars/binaries/triples etc. to estimate the size and properties of the population of merging BH/NS binaries that can be compared with GW observations. 
Note that the formation of stars in some specific environments relevant for the formation of DCO mergers (e.g. globular clusters) may not follow the overall star formation history of the Universe.
To model the population of DCO mergers formed in such environments, metallicity-dependent star formation history appropriate for those systems needs to be considered.
    }             
    \label{fig: schematic population modelling}
%   \end{minipage}
\end{figure}

Modelling the population of DCO mergers that may exist in the real Universe 
consists of two main, independent steps (see Figure \ref{fig: schematic population modelling}).
First, one has to describe how many stars in the considered environment form at different times and what are their birth metallicities.
This provides initial conditions for the second part of the population modelling, which describes the evolution of those stars.
In this step one has to account for any potential interactions with other stars/compact objects (e.g.\ mass transfer phases in binaries/multiples and any relevant dynamical effects) and the surrounding medium (e.g.\ when the evolution takes place in a dense AGN disk).
The combination of those two parts allows one to obtain the model population of GW sources, describing the rate and properties of DCO mergers as a function of time/redshift.
\\
Different merger formation channels (i.e.\ different ways to answer the question: how to bring two BH/NS sufficiently close together so that they merge within a time shorter than the Hubble time) were proposed in the literature
(see e.g.\ \cite{Barack19} for an overview).
Various groups often describe the formation of DCO mergers focusing on a particular class of formation channels (e.g.\ considering only isolated stellar binaries, where interactions with other stars/compact objects/their environment can be neglected; but see \cite{Mapelli22}).
Many formation paths may contribute to the observed population.
In particular, in the case of BH-BH mergers comparable local merger rate densities were reported for different formation channels (see e.g.\ the recent compilation by  \cite{MandelBroekgaarden21}).
To consistently model the population of DCO mergers combining contributions from different formation scenarios it is necessary to know how the cosmic star formation is distributed between the different environments.
The focus of this review is on the DCO mergers formed in environments in which the star formation follows the overall cosmic star formation history.
This may not be the case for globular/nuclear clusters and AGN disks (see Section \ref{subsection: other channels}).
\\
Each of the proposed formation channels allows one to predict distributions of various properties (e.g.\ masses)
of DCO mergers produced from a coeval, homogeneous population of progenitor stars.
One of the characteristics obtained is the distribution of delay times.
Note that even if the predicted delay-time distribution strongly favors short merger times
(as commonly found in the isolated binary evolution channels, e.g.\ \cite{Belczynski16}), 
it does not imply that the fraction of local mergers originating from the early Universe is negligible.
Firstly, their contribution to the observed population is amplified by the cosmic star formation history: the star formation rate density used to be higher in the past ($>$10 times higher 10 Gyrs ago than at present \cite{MadauDickinson14,Gruppioni20,Chruslinska21}).
Therefore, also BH and NS were forming at a higher rate.
Secondly, the abundance of elements heavier than helium (metallicity) with which the stars are forming changes over time, shifting towards higher values as the next generations of stars evolve and pollute the surrounding medium with enriched material \cite{Burbridge57,Nomoto13,MaiolinoMannucci19}.
This is important, because stellar evolution (and the outcome of this evolution) is a metallicity dependent process (see Section \ref{subsec: metallicity and stellar evolution}).
As a consequence, the efficiency of formation of merging DCO (i.e.\ the number of systems formed per unit of star formation) and the properties of mergers also depend on metallicity \cite{Belczynski10a,Dominik13,MennekensVanbeveren14,Stevenson17,Giacobbo18,Klencki18,Kruckow18,ChruslinskaNelemansBelczynski19,EldridgeStanwayTang19,Santoliquido21,Broekgaarden22}.
Therefore, the modelling (and interpretation) of the properties of the population of any stellar-evolution related transients, in particular GW sources, requires the knowledge of not only the total star formation rate density (SFRD), but also its metallicity distribution  - i.e.\ the metallicity dependent cosmic star formation history (\cite{ChruslinskaNelemans19}; see the left part in Figure \ref{fig: schematic population modelling}).

\subsection{Birth metallicity and the evolution of massive stars}\label{subsec: metallicity and stellar evolution}

\subsubsection{Note on the definitions of metallicity}\label{sec: metallicity note}

The term ‘metallicity‘ is generally understood as a measure of the abundance of metals (elements heavier than helium) in a certain object or system.
However, different definitions are used in the literature.
In theoretical studies metallicity is typically defined as a fraction of total baryonic mass contained in metals:
$\rm Z = M_{metals}/M_{baryons}$.
Observationally, it is often more convenient to express this quantity in terms of relative abundances of particular elements.
The most commonly used measures rely on the
oxygen to hydrogen abundance ratio:
$\rm Z_{O/H} = 12 + log_{10}(O/H) = 12 + log_{10}(n_{O}/n_{H})$,
or iron to hydrogen abundance ratio relative to solar:
$\rm [Fe/H]=log_{10}(n_{Fe}/n_{H}) - log_{10}(n_{Fe \odot}/n_{H \odot})$.
 In those expressions, n$_{i}$ stands for the number density of either oxygen, iron or hydrogen as indicated by the index and the symbol $\odot$ refers to the reference solar values.
 It is often assumed that the other elements scale linearly with the ones measured maintaining the solar abundance ratios (e.g. $\rm log_{10}(Z/Z_{\odot} ) =
log_{10}(n_{O}/n_{H}) - log_{10}(n_{O\odot}/n_{H\odot})$).
However, in some environments (such as young, highly star forming galaxies; see Section \ref{subsec: major issues}) this is generally not the case and non-solar composition with $\alpha$-enhanced ratios (e.g. $\rm [O/Fe]=log_{10}(n_{O}/n_{Fe}) - log_{10}(n_{O \odot}/n_{Fe \odot})>$0) is often found.\\
Whenever it is relevant to refer to a particular definition of metallicity, it is explicitly mentioned in the text.
If specific choice of solar metallicity is needed, the values reported by \cite{GrevesseSauval98} are used (i.e.\ solar metal mass fraction Z$_{\odot}$=0.017 and solar oxygen abundance $\rm Z_{O/H \odot}$=12 + log$\rm _{10}(n_{O \odot}/n_{H \odot}$)=8.83).

\subsubsection{Single stars}
Birth metallicity is one of the most important properties that determine the evolution and fate of a massive star \cite{Maeder92,Hurley00,Heger03,Langer12}.
The most striking example of its impact on evolution of single stars is related to the amount of mass lost in stellar winds \cite{Kudritzki87,Vink01,Mokiem07,GrafenerHamann08,Puls08,VinkSander21}.
Winds of hot, massive stars are driven by the radiation
pressure on metal lines and are primarily sensitive to the abundance of iron (which easily dominates the atmospheric opacity due to its complex atomic structure, e.g. \cite{VinkDeKoter05,VandenBerg12}). Mass loss is thus much more severe at solar-like metallicity than in low metallicity (iron-poor) environments.
This affects the mass of the star and the final remnant, leading to a
metallicity-dependent maximum BH mass that can result from single star evolution (limited to only $\sim$ 20 M$_{\odot}$ at solar metallicity, e.g.\ \cite{Maeder92,FryerKalogera01,EldridgeTout04,Belczynski10b,Fryer12} as also reflected in the masses of BHs in nearby X-ray binaries \cite{Tetarenko16}).
Finally, reduced wind mass loss at low metallicity 
helps to retain angular momentum within the star
that is born rapidly spinning, possibly allowing for the
formation of a rapidly spinning NS/BH \cite{MaederMeynet00} and chemically homogeneous evolution of the progenitor star \cite{Maeder87,deMink09}.

\subsubsection{Stellar binaries and higher-order multiples}

Additional metallicity-dependent effects can non-straightforwardly affect the evolution of stars in binaries/multiples.
Mass and angular momentum escaping the system due to stellar winds affects the orbital separation (typically widening the orbit \cite{Schroder21}, i.e.\ leading to wider binaries at higher metallicity).
Furthermore, evolution of stellar radii, in particular the degree of post-main sequence radial expansion depends on metallicity \cite{BrunishTruran82,BaraffeElEid91}.
This affects when/if the evolving star overfills its Roche lobe (and starts transferring mass to its companion) and therefore affects the nature and outcome of stellar interactions, as well as the observed characteristics of the system (e.g.\ \cite{Vanbeveren98,Hurley00,deMink08,Eldridge17,Klencki20,Laplace20,Bavera21,Klencki21b}).
Both winds and binary interactions can affect the mass and structure of stellar envelope and the final core \cite{Woosley2019,Laplace21,Schneider2021} in a way that depends on metallicity \cite{Klencki21b,AguileraDena2022}.
Stars stripped of their envelopes eject less mass from the system during the core-collapse/supernova and potentially lead to the formation of NS/BH with small velocities (so-called natal kicks, \cite{Janka17}).
This increases the chances that the binary/multiple remains bound after the formation of the compact object.
Mass loss (and possibly natal kick) can be reduced  for BH progenitors with sufficiently massive cores that can develop in low metallicity stars \cite{JankaMueller94,FryerKalogera01,Fryer12,Chan20}.
Such progenitors are predicted to lead to weak supernovae (with significant fallback accretion onto the newly born compact object, reducing the kick and ejecta) or, in case of the most massive progenitors - collapse to BH directly with no/negligible kick and mass ejection \cite{Fryer99,Smartt09}.
This supports the formation of binaries with (massive) BH at low metallicity \cite{Belczynski10a,Mapelli10}.
Finally, the fraction of stars that are born in close binaries/multiples may itself depend on metallicity. Some evidence of such dependence has been reported for lower mass stars (i.e.\ white dwarf progenitors) suggesting higher fraction of multiples at lower metallicity \cite{Moe19}. A fixed fraction is typically assumed when modelling the population of massive stars in binaries (but note that this fraction is already very high, with typical assumptions ranging between 0.7--1).

\subsubsection{Stars in dense environments}\label{subsec: dense environs}

Through its impact on the masses and birth velocities of compact objects, birth metallicity may have an indirect effect on the evolution of stars in dense environments (open / globular/ nuclear clusters, AGN disks).
More massive stars/compact objects (originating from low metallicity progenitors) are more likely to sink deeper in the potential well of their host system, occupying the regions where dynamical encounters can happen frequently (e.g. \cite{PortegiesZwart10}). By receiving weak/no natal kick, they can preferentially remain within those regions.
The above effects lead to relatively low expected contribution of formation channels involving dynamical effects in dense environments to the formation of DCO mergers involving NS (e.g.\ \cite{Belczynski18,Zevin19,Ye20}).

\section{The sensitivity of GW sources to metallicity-dependent cosmic star formation history}

There is now a substantial body of literature showing that different assumptions about the metallicity-dependent cosmic star formation history (called f$_{\rm SFR}$(Z,z) in the reminder of this paper) can significantly affect the modelled properties of the population of DCO mergers (e.g.\ \cite{ChruslinskaNelemansBelczynski19,Neijssel19,Tang20,Briel21,Santoliquido21,Broekgaarden22}).
The choice of the f$_{\rm SFR}$(Z,z) affects nearly all observable
properties of the population of DCO mergers:
their merger rate density, the relative rate of different types of mergers (e.g.\ the expected ratio of the rate of NS+NS to BH+BH mergers) and their mass distribution.
Somewhat less clear is the potential effect of the assumed f$_{\rm SFR}$(Z,z) on the (effective) spin distribution of DCO mergers (as could be expected if, for instance, the spin magnitude would strongly depend on the mass -- and hence indirectly on metallicity -- of the progenitor star).
Furthermore, f$_{\rm SFR}$(Z,z) affects the evolution of merger properties as a function of time/redshift. 
While the most distant event observed with the current network of detectors comes from redshift$\sim$0.8 \cite{GW190521,GWTC3}, future generation of GW detectors will allow to
map the properties of DCO mergers up to high redshifts (beyond redshift$>$10 for BH+BH mergers \cite{Punturo10,Reitze19,Maggiore20,BorhanianSathyaprakash22}).
Modelling and understanding those redshift trends is thus particularly interesting in the context of future detectors.\\
So far, the impact of f$_{\rm SFR}$(Z,z) on the population properties of GW sources has been mostly discussed for DCO mergers formed through isolated binary evolution and that is necessarily the focus of this review.
I briefly comment on other DCO merger formation channels in Section \ref{subsection: other channels}.

\subsection{DCO mergers formed in isolated binary evolution}

Merging DCO formed in the most common variations of the isolated channel originate from progenitor binaries sufficiently wide to avoid premature merger (when the progenitor stars evolve and expand) but sufficiently narrow to ensue interaction(s) through mass transfer phase(s). Mass transfer (either stable or unstable - leading to a common envelope evolution \cite{Ivanova13,Ivanova20}) is key to shrinking the orbit to the separation required for the DCO to merge within the Hubble time \cite{Paczynski76,vandenHeuvel76,MennekensVanbeveren14, Belczynski16,EldridgeStanway16,Stevenson17,Klencki18,MapelliGiacobbo18,Kruckow18,MapelliGiacobbo18,Breivik20,Olejak21}. Under special circumstances (in particular low metallicity and high masses of the progenitors), stars may evolve chemically homogeneously and avoid strong radial expansion, allowing to produce merging DCO from initially narrow binaries \cite{Cantiello07,deMinkMandel16,MandeldeMink16,Marchant16}.
Models of the populations of DCO mergers originating from isolated binaries are long known to suffer from large uncertainties due to poorly understood phases of evolution of massive stars in binaries (in particular binary interactions; see \cite{MandelBroekgaarden21} and references therein for recent examples).
More recently, it became apparent that the assumptions about the f$_{\rm SFR}$(Z,z) can lead to comparable uncertainty in the model predictions (e.g. \cite{ChruslinskaNelemansBelczynski19,Neijssel19,Tang20,Boco21,Santoliquido21,Broekgaarden21,Briel21,Broekgaarden22}).
Interestingly, f$_{\rm SFR}$(Z,z) is found to have the strongest effect on the properties of the population of BH+BH mergers. 
The effect is weaker in case of BH+NS mergers and the properties of NS+NS mergers appear to be relatively mildly affected by f$_{\rm SFR}$(Z,z) (when compared to the impact of different assumptions about the binary evolution, e.g.  \cite{ChruslinskaNelemansBelczynski19,Santoliquido21,Broekgaarden22}).
To understand this behaviour, it is instructive to look at the efficiency of formation of various types of DCO mergers as a function of metallicity.
Such dependence can be inferred theoretically with binary population synthesis simulations.
  \begin{figure}[h!]
  \begin{minipage}[c]{0.55\textwidth}
    \includegraphics[width=1.\textwidth]{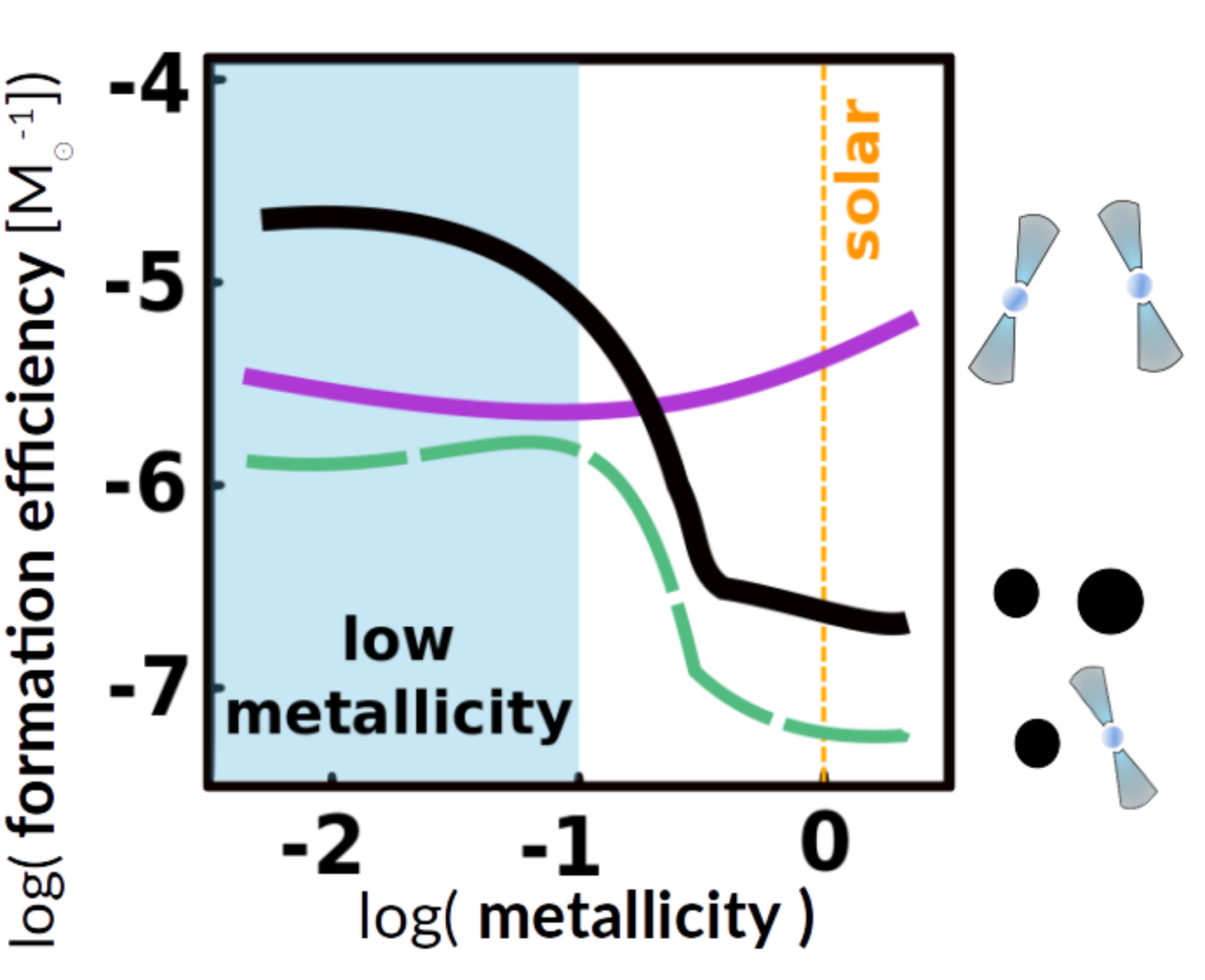}
  \end{minipage}\hfill
  \begin{minipage}[c]{0.45\textwidth}
    \caption{
Sketch illustrating the commonly found dependence of the efficiency of formation of merging DCO (black - BH+BH, green - BH+NS, purple - NS+NS) as a function of metallicity (see Section \ref{subsec: simple Xeff}).
Note the strong decrease in the number of merging DCO containing BH formed per unit of stellar mass above a certain metallicity. The exact characteristics of this dependence are sensitive to the detailed description of the evolution of massive stars in binaries (see Section \ref{subsec: evolution effects on Xeff}).
    }             
    \label{fig: schematic Xieff}
  \end{minipage}
\end{figure}

Note that how exactly the efficiency of formation of DCO mergers varies with metallicity depends on the binary population synthesis model (i.e.\ the particular set of assumptions used in the simulations to parameterize the physics of binary interactions, stellar winds, core collapse physics and initial binary parameters; see Section \ref{subsec: evolution effects on Xeff}. ).
Nonetheless, the relation shows some characteristic features that are present across most of the population synthesis models considered in the literature.
The simplified version of this dependence, which captures those common features, is illustrated in Figure \ref{fig: schematic Xieff} and is discussed in Section \ref{subsec: simple Xeff}. 

\subsubsection{The `typical' dependence on metallicity}\label{subsec: simple Xeff}

The most striking feature shown in Figure \ref{fig: schematic Xieff} is the strong decrease (even by a few orders of magnitude) in the number of merging BH+BH binaries produced per unit of mass formed in stars that appears above a certain metallicity.
In other words, BH+BH mergers are thought to form much more efficiently at low metallicity than in solar-like metallicity environments (as indicated already in e.g.\ \cite{Belczynski10a,Mapelli10}).
Compared to the BH+BH dependence, the formation efficiency of merging NS+NS barely depends on metallicity. 
The corresponding relation for merging BH+NS systems shows a similar decrease in the efficiency of formation towards high metallicities as in case of BH+BH binaries, albeit weaker. 
All relations flatten at low metallicity.\\
Those trends are the result of many complex processes involved in binary/stellar evolution acting together \cite{Giacobbo18,ChruslinskaNelemansBelczynski19,Santoliquido21,Broekgaarden22}.
However, qualitatively they can be understood by considering the effects that are the most straightforwardly linked to metallicity (see Section \ref{subsec: metallicity and stellar evolution}).
Weaker stellar winds at lower metallicities lead to tighter binaries with more massive compact objects. Under the typical assumptions, more massive BHs receive smaller natal kicks, which increases the fraction of binaries that remain bound after the supernova.
Below a certain metallicity, BH formation with no natal kick and mass ejection becomes common in most models.
Finally, metallicity dependence of the radial expansion causes more stars to engage in mass transfer early in their evolution (before they reach the giant phase) at high metallicities. Such interactions are thought to be more likely to lead to binary merger before a DCO is formed.
All of these effects support the formation of merging DCO containing a BH at low metallicity.
NS originate from stars with masses within a much narrower range than BH (their evolution is more alike) and are limited to relatively low mass progenitors (with initial masses $\sim$10 M$_{\odot}$ in merging NS+NS, e.g. \cite{Tauris17,Chruslinska18,VignaGomez18}).
The impact of winds on the evolution of such stars is much weaker, which is reflected in the relatively weak metallicity dependence of the efficiency of formation of NS+NS mergers.

\subsubsection{The impact of evolutionary assumptions on the DCO merger formation efficiency} \label{subsec: evolution effects on Xeff}

While the qualitative behaviour of the DCO formation efficiency as a function of metallicity discussed in Section \ref{subsec: simple Xeff} is found across many variations of input assumptions describing the evolution of massive stars in binaries, the exact characteristics of this dependence differ (see examples in e.g.\ \cite{ChruslinskaNelemansBelczynski19,Santoliquido21,Broekgaarden22}).
Note that the apparent consensus about the very strong low metallicity preference of BH+BH and metallicity insensitivity of NS+NS mergers might (to some extent) result from similar assumptions used by different groups.
The huge parameter space involved in the simulations makes it difficult to fully assess the robustness of those trends. 
Most of the current predictions rely on results obtained with rapid codes, utilizing simplified prescriptions to describe complex phases of stellar/binary evolution (
notable exceptions capturing stellar evolution and the physics of binary interactions in more detail while still allowing to model representative binary/transient populations include BPASS \cite{Eldridge17, StanwayEldridge18} and the recently released POSYDON \cite{Fragos22} codes).
Calculations performed with detailed codes indicate that some relevant effects may be missed by those approximate recipes
(for instance, important differences were found in the evolution of stars stripped of their envelopes by binary interactions \cite{Laplace20}, mass transfer stability and common envelope ejection criteria \cite{Klencki21a,Gallegos-Garcia21,Marchant21}).
Furthermore, even the rapid population synthesis studies typically explore models varying only one major parameter at a time.
\\
Strikingly, the explored evolutionary assumptions mostly affect the shape of the high metallicity part of the relation and the location and steepness of the drop, while the low metallicity part remains relatively flat and featureless.
This reflects the fact that within the current models, there are no strongly metallicity-dependent processes that become efficient at very low metallicities (where the impact of winds is negligible and the contribution of BH formed in direct collapse to the population of DCO mergers saturates).
However, the value at which the relation flattens is model dependent (especially for NS+NS mergers). This value can be easily influenced, for instance, by varying assumptions that affect the efficiency of the orbital tightening during the late mass transfer phase(s).
 Note that while the fact that the winds of massive stars depend on metallicity is well established, the exact relation is uncertain. 
Given the arguments quoted in Section \ref{subsec: simple Xeff}, one can expect that the assumed relation affects the steepness and location of the drop in the efficiency of formation of DCO with BH (e.g.\ if the winds become weaker at higher metallicity, the characteristic drop also shifts to the higher metallicity). 
Such behaviour is indeed observed in \cite{Broekgaarden22} (compare their model S and T).
The magnitude of the drop can be reduced, for instance, if the BH natal kick is less tightly related to progenitor mass (and so metallicity) than typically assumed.
Finally, detailed stellar evolution models repeatedly show that the envelope binding energy and mass transfer stability (properties that are crucial for the description of binary interactions) non-trivially depend on the structure of the star and its envelope properties (e.g.\ \cite{PavlovskiiIvanova15,Kruckow16,Pavlovskii17,Klencki21a,Marchant21}).
In combination with metallicity-dependent evolution of stellar radii and the so far poorly explored impact of metallicity on stable mass-transfer evolution \cite{Laplace20, Klencki21b}, this may induce additional metallicity effects on the formation efficiency that are not fully accounted for in current population synthesis simulations.
\\
Note that (some degree of) low metallicity preference for BH+BH merger formation can be expected to hold in different variations of isolated binary/stellar multiple evolution (also those that are not explicitly accounted for in the studies discussing the DCO formation efficiency referenced throughout this section), since all of them are affected by the main factors discussed in Section \ref{subsec: simple Xeff} (i.e.\ stellar winds, supernovae kick/ejecta). 
Additionally, so-called chemically homogeneous evolution of isolated binaries (usually considered a separate BH+BH merger formation channel \cite{Cantiello07,deMinkMandel16,MandeldeMink16,Marchant16}) is expected to operate exclusively at low metallicity and only for relatively massive BH progenitors.
However, metallicity dependence of dynamically formed DCO mergers may significantly deviate from the trends discussed in this Section (see Section \ref{subsection: other channels}).

\subsubsection{The effect of the metallicity-dependent cosmic star formation history on the population properties}

It is worth highlighting a few consequences of the metallicity dependence of DCO mergers discussed in Sections \ref{subsec: simple Xeff} and \ref{subsec: evolution effects on Xeff}.
First, the quantitative effect of the assumed f$_{\rm SFR}$(Z,z) on the population properties depends on the adopted evolution/DCO formation model: if the model leads to weak metallicity dependence (of BH/NS mass, formation efficiency), the assumed birth metallicity distribution of stars has relatively small impact on the population properties (compared to evolutionary assumptions).
This appears to be the case for the current NS+NS merger formation models. Note that the uncertainty in the total SFRD (which is substantial at high redshifts, see e.g.\ section 6.1 in \cite{Chruslinska21}) can still add non-negligible uncertainty to the predicted population properties of those systems (especially if redshift trends are considered).
On the other hand, the strong low metallicity preference of BH+BH mergers suggested by current models leads to the conclusion that their population is relatively insensitive to many assumptions about the evolution of massive stars \cite{Santoliquido21,Broekgaarden22}.
This is true for a range of assumptions used in current models that mostly affect the evolution at high metallicity (but see the caveats discussed in Section \ref{subsec: evolution effects on Xeff}), where the formation efficiency of BH+BH mergers is order(s) of magnitude lower. 
In such cases, their population properties are shaped by the low metallicity part of the cosmic star formation history (SFH, see Section \ref{subsec: BHBH mergers}).
\\
Qualitatively, one can expect that assuming a f$_{\rm SFR}$(Z,z) with a higher fraction of recent low metallicity star formation leads to a DCO merger mass distribution containing a higher fraction of massive events.
Such a f$_{\rm SFR}$(Z,z) also leads to a higher local rate of BH+BH and (with a smaller increase) BH+NS mergers, affecting the relative ratio of rates of different systems. Note that such behaviour still leaves multiple routes for the interpretation of the observed population of GW sources.
In principle, a higher than expected fraction of massive BHs among the observed events can be attributed to (i) a higher fraction of star formation happening at low metallicity, (ii) weaker stellar winds of massive stars, (iii) a higher fraction of BH mergers forming with long delay times or (iv) sequential BH mergers in dense environments (to list a few possibilities).
Identifying ways to break such degeneracies is one of the major challenges in GW astrophysics.

\subsection{Consequences of the low metallicity preference of BH+BH merger progenitors}\label{subsec: BHBH mergers}

\begin{figure}[h!]
  \begin{minipage}[c]{0.55\textwidth}
    \includegraphics[width=1.\textwidth]{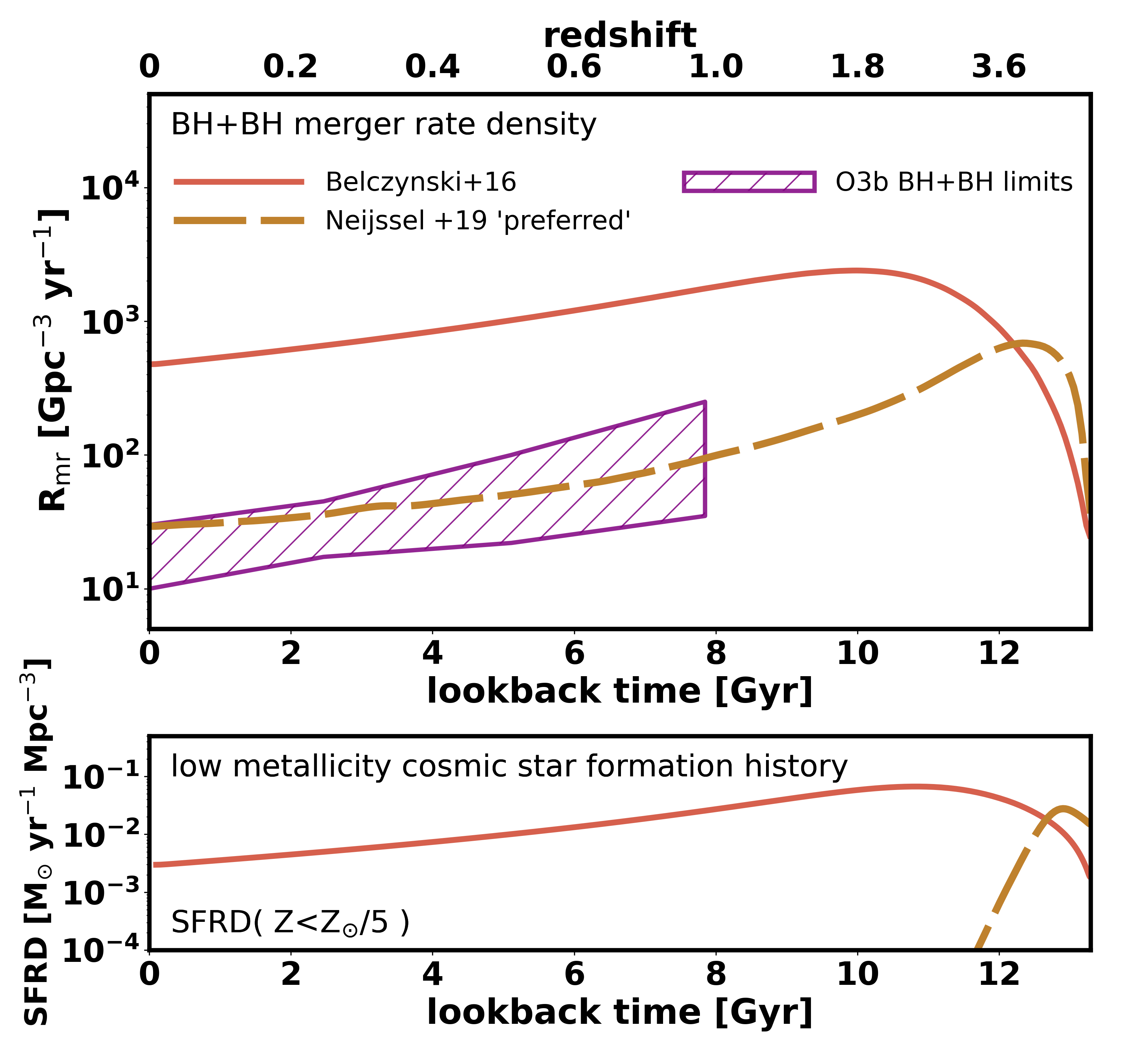}
  \end{minipage}\hfill
  \begin{minipage}[c]{0.45\textwidth}
    \caption{
    Top: The BH+BH merger rate density as a function of lookback time/redshift shown for one example population synthesis model (fiducial from Broekgaarden et al.\ 2022) and two (very) different models of metallicity-dependent cosmic SFH used in the literature (solid red - as used in Belczynski et al.\ 2016, dashed brown - `preferred' model from Neijssel et al.\ 2019; see bottom panel). The hatched region indicates current GW-based constraints (\cite{GWTC3pop}, accounting for the rate evolution with redshift).\newline
    Bottom: The star formation rate density (SFRD) that falls below 1/5 solar metallicity for the two models of the metallicity-dependent cosmic SFH used in the top panel.
    }             
    \label{fig: rate example}
  \end{minipage}
\end{figure}

\begin{figure}[h!]
  \begin{minipage}[c]{0.55\textwidth}
    \includegraphics[width=1.\textwidth]{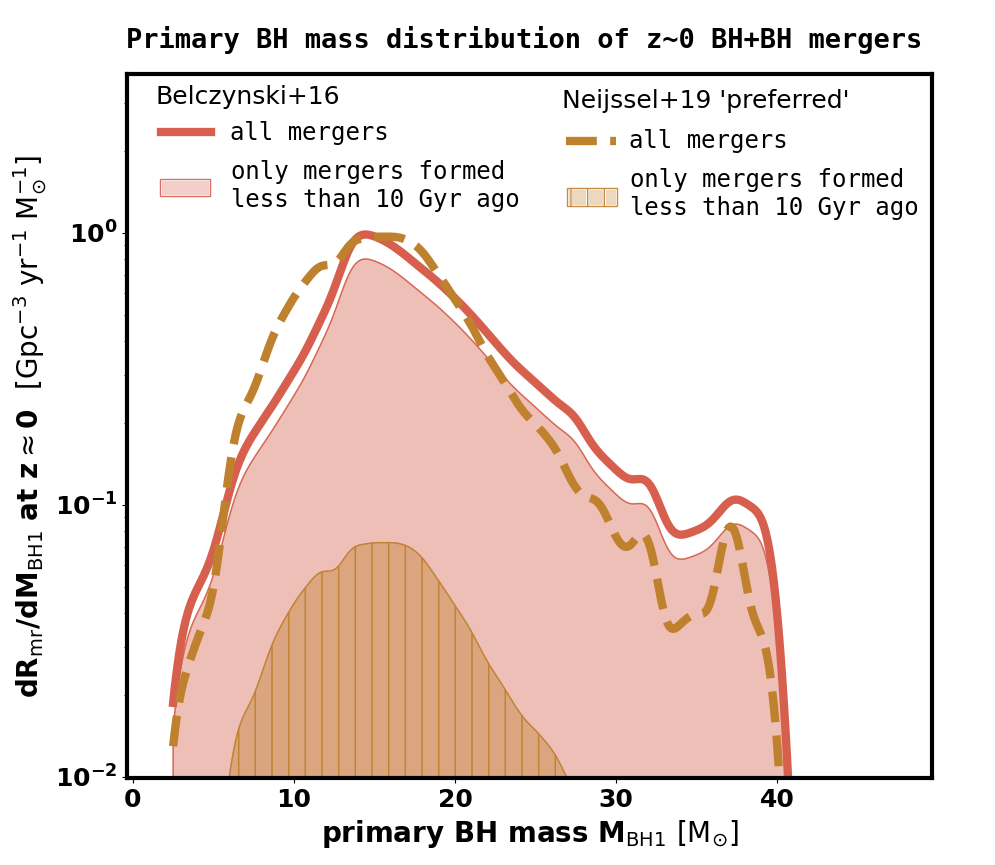}
  \end{minipage}\hfill
  \begin{minipage}[c]{0.45\textwidth}
    \caption{Local BH+BH merger rate density split in bins of primary (more massive) BH mass M$_{\rm BH1}$ in the merging systems for the same population synthesis model and two (very) different models of metallicity-dependent cosmic SFH as used in Figure \ref{fig: rate example}.
    Distributions were normalized by the peak rate value to facilitate comparison.
    Filled parts of the distributions correspond to local BH+BH mergers that originate from progenitors formed less than 10 Gyr ago (at redshift $<$2).
    If the low metallicity SFR is limited to very high redshifts (model with hatched area and dashed line), the vast majority of the predicted local BH+BH mergers (and nearly all massive BH $>$30 M$_{\odot}$) originate from stars formed more than 10 Gyr ago (i.e.\ beyond the peak of the cosmic SFH).
    }             
    \label{fig: mass example}
  \end{minipage}
\end{figure}

To illustrate the consequences of the strong metallicity dependence of BH+BH mergers I consider a simple example: a single model of binary evolution is combined with two very different literature assumptions about f$_{\rm SFR}$(Z,z) in order to compare the resulting properties of the population of BH+BH mergers (merger rate density - Figure \ref{fig: rate example} and mass distribution - Figure \ref{fig: mass example}). All figures show the intrinsic properties of the population (i.e.\ not taking into account the sensitivity of GW detectors which is needed to model the mock `observed' population).
All models and tools used in this Section are publicly available and are described in detail in the following references:
Broekgaarden et al.\ (2021) \cite{Broekgaarden22, COMPAS:2021methodsPaper} -
isolated binary population synthesis model (model A/fiducial), showing the typical metallicity dependence described in Section \ref{subsec: simple Xeff}, Belczynski et al. (2016) \cite{Belczynski16} -
f$_{\rm SFR}$(Z,z) with a significant fraction of star formation at low metallicity even at low redshifts, Neijssel et al.\ (2019) \cite{Neijssel19} - their `preferred' f$_{\rm SFR}$(Z,z) model, characterised by a negligible amount of recent low metallicity star formation.
The low metallicity cosmic SFH (here chosen as the SFRD happening below 1/5 solar metallicity) resulting from the two f$_{\rm SFR}$(Z,z) assumptions are contrasted in the bottom panel of Figure \ref{fig: rate example}.
The adopted 1/5 solar metallicity threshold was chosen to select the star formation happening at metallicities below the abrupt drop seen in the metallicity dependence of the formation efficiency of BH+BH mergers for the evolutionary model considered in this example. Therefore, it allows to zoom into the SFH at metallicities that dominate the formation of merging BH+BH (the location of this drop, defining the low metallicity regime of BH+BH merger formation is model dependent; see also Section 5 in \cite{Chruslinska21} for further discussion of the low metallicity SFH threshold).
\\

The considered example highlights the importance of the assumed f$_{\rm SFR}$(Z,z)
for the interpretation of the observed properties of BH+BH mergers:
\begin{itemize}
    \item f$_{\rm SFR}$(Z,z) strongly affects both the shape and the normalization of the BH+BH merger rate density - neither can be straightforwardly linked to assumptions about the evolution/BH+BH formation.
    Depending on the assumed f$_{\rm SFR}$(Z,z) one can reach different conclusions about the binary evolution when the modelled rate is compared with GW observations (see the top panel in Figure \ref{fig: rate example}).
    Similarly, not taking into account the differences in the assumed f$_{\rm SFR}$(Z,z) can lead to erroneous conclusions when the population models delivered by different groups are compared. See Section \ref{subsec: degeneracies and 3G} for further discussion.
    \item The modelled redshift dependence is shaped by the assumed low metallicity cosmic SFH (compare the top and bottom panels of Figure \ref{fig: rate example}).
    This indicates the possibility to probe the metallicity-dependent cosmic SFH with GW observations (see Section \ref{subsec: GW to study galaxies}).
    \item f$_{\rm SFR}$(Z,z) affects mostly the high mass end of the GW-observed BH mass distribution (see Figure \ref{fig: mass example}). Note that BH mergers formed dynamically are also expected to affect primarily this part of the mass distribution. This again adds to challenges in the interpretation of GW observations.
    \item Depending on the assumed f$_{\rm SFR}$(Z,z), one can reach drastically different conclusions about the origin of the locally observed population. 
    Assuming a f$_{\rm SFR}$(Z,z) with a negligible amount of recent low metallicity star formation leads to an expectation that the vast majority of local BH+BH mergers (and nearly all massive events) come from progenitor stars formed in the early Universe (compare thick and thin lines in Figure \ref{fig: mass example}, the latter selecting only local mergers formed less than 10 Gyr ago).    Consequently, in this case the host galaxies of BH+BH mergers could show very different properties than the galaxies in which their progenitors formed. 
    This is particularly important when ranking the galaxies during searches for potential electromagnetic counterparts to GW events (see Section \ref{subsec: hosts})
\end{itemize}
\subsubsection{Impact on merger host galaxy properties}\label{subsec: hosts}

Galaxies evolve due to their own star formation, gas inflows/outflows, mergers and other interactions with nearby galaxies. 
In the considered example (Figure \ref{fig: mass example}), if f$_{\rm SFR}$(Z,z) with low metallicity star formation concentrated at high redshifts is assumed, a significant fraction of the local BH+BH merger progenitors forms in young, active galaxies $>$10 Gyr ago. 
Their mergers can generally be expected to happen in evolved, massive hosts with little ongoing star formation.
Conversely, in case of f$_{\rm SFR}$(Z,z) allowing for a high level of low metallicity star formation even at low redshifts, more mergers would be expected in galaxies that are currently producing metal poor stars (relatively low mass and highly star forming).\\
To date, the properties of merger host galaxies were studied predominantly with the use of large-scale cosmological simulations (e.g. \cite{Artale19,Toffano19,Mandhai21,Chu22}). Simulations allow to directly connect the progenitor formation galaxy with the merger host, accounting for the star formation and merger history of the galaxy that happened in the meantime. This is generally not possible in the observation-based approach.
However, while such simulations provide valuable insights, they do not show as much variation in the metallicity-dependent cosmic star formation history as spanned by the observation-based estimates and struggle to resolve low mass galaxies (see Section \ref{sec: fSFR(Z,z)}).
Based solely on their results one may thus underestimate the uncertainty and the range of likely merger host properties. Future studies should address this question and discuss its importance for the searches for electromagnetic counterparts to GW events.
Important steps in this direction were recently taken by \cite{Boco20,Boco21} (combining observation-based f$_{\rm SFR}$(Z,z) with crude star formation histories of typical star forming galaxies at different redshifts) and \cite{Santoliquido22} (proposing a method to combine observation based f$_{\rm SFR}$(Z,z) with merger trees from cosmological simulations).

\subsection{DCO mergers formed in other formation channels}\label{subsection: other channels}

To self-consistently model the contribution of all of the proposed DCO merger formation scenarios to the final population, one also needs to know what fraction of the total metallicity-dependent cosmic star formation history is 
used to form stars in different environments: i.e.\ the galactic field (where dynamical interactions are negligible and binary/stellar multiples evolve undisturbed) and dense environments such as young/open/globular clusters, nuclear clusters (the latter potentially also affected by the presence of AGN disk at some point during its life, which may enhance BH+BH merger production, e.g.\ \cite{Bartos17,McKernan18,Tagawa20}).
Furthermore, one needs to know the distribution of initial parameters of clusters (such as mass, density - which affect the subsequent evolution of the system) and has to be able to specify what fraction of stars forming in dense environments dissolves into field before any dynamical effects can affect their evolution.
The history of the formation of globular/nuclear clusters is debated, with different scenarios proposed in the literature \cite{West04,MuratovGnedin10,Kruijssen15,RodriguezLoeb18,ElBaldry19}. Observations of local globular clusters reveal that their stellar populations are old and metal-poor, suggesting that their star formation was completed at high redshift \cite{Harris96}. In that sense, even if globular clusters were to dominate the production of BH+BH mergers, one would still expect some connection between the formation of those mergers and the low metallicity cosmic star formation history.
This link is less clear in case of nuclear clusters, which may form from dissolving globular clusters, stars accreted from infalling galaxies, in-situ star formation or some combination of those \cite{Neumayer20}.
On the other hand, star formation in young stellar clusters is expected to follow the cosmic SFH. Recent studies suggest that the fraction of the total star formation happening in such environments amounts to $\sim$10\% (but much higher values are also quoted in the literature - see \cite{Mapelli22} and references therein).
Interestingly, \cite{DiCarlo20} show that the metallicity dependence of the dynamically assembled BH+BH mergers in young star clusters can be significantly flatter than in the field.
However, a significant fraction of merging BH+BH formed in such (relatively sparse) clusters may originate from effectively isolated binaries. 
Depending on their relative contribution
(which likely varies with the cluster properties \cite{Torniamenti22}), the overall BH+BH merger formation efficiency in young clusters may still show a prominent decrease towards high metallicity.
Overall, the (redshift-dependent) properties of the DCO merger population predicted by the cluster scenarios strongly depend on the assumed cosmic history of formation of those stellar systems  \cite{Romero-Shaw21, Mapelli22}.

\subsection{Gravitational waves as a tool to study chemical evolution of galaxies} \label{subsec: GW to study galaxies}
The fact that the population properties of DCO mergers depend on f$_{\rm SFR}$(Z,z) can in principle be used to learn about the chemical evolution of galaxies.
Different types of DCO mergers appear to be sensitive to different parts of f$_{\rm SFR}$(Z,z): NS+NS mergers might better probe the total SFRD, while BH+BH the SFRD happening at low metallicity.
The predicted strong low metallicity preference of BH+BH mergers is particularly interesting in this context.
If correct, it can help to understand the properties of those galaxies that are the most difficult to constrain with electromagnetic observations (faint, high-redshift objects; see also \cite{Vitale19,Graziani20}).
Future GW detectors will map the properties of BH+BH mergers as a function of redshift \cite{Punturo10,Reitze19,Maggiore20}, potentially allowing to constrain the overall low metallicity cosmic SFH (see the example shown in Section \ref{subsec: BHBH mergers}). This in turn 
can be linked, for instance, to the
slope of the faint end of the galaxy luminosity/mass function or properties of starburst galaxies (both strongly affect rate at which the low metallicity SFRD increases with redshift and its peak location, see  \cite{Chruslinska21}).
It is an interesting possibility to consider in future studies exploring the science capabilities of the third generation GW detectors.
\\
Furthermore, to model the populations of GW sources, it is necessary to account for stars forming in environments significantly different from the current Milky Way, in which the stellar initial mass function (IMF) may be different \cite{Hopkins18}.
When the IMF and the star formation rate are varied consistently, the overall effect on the formation of BH progenitors is negligible (but it may still leave imprints on the observed mass distribution of BH mergers; \cite{Chruslinska20}).
However, IMF variations may noticeably affect the formation of intermediate mass white dwarf/NS progenitors \cite{Chruslinska20} - the proposed space-based GW detectors will be sensitive to close binaries with such compact objects in the nearby Universe \cite{LISA}.
Given the large uncertainties involved in the models of DCO merger formation and the IMF variations, it is difficult to speculate whether such IMF-related effects could be observed with future GW detectors.
\\
Finally, future GW observations can potentially be used to study the cosmic history of globular/nuclear cluster formation \cite{Romero-Shaw21}.
\\
The feasibility of many of the science goals of future ground-based GW detectors heavily depends on our ability to distinguish sub-populations of DCO mergers formed in different channels \cite{Ng21} (especially for BH+BH mergers, for which the contribution of different channels may be comparable and strongly vary with redshift).
In particular, understanding of the redshift trends (e.g.\ the location of any potential peaks/features in the redshift-dependent BH+BH merger rate/mass distribution) of the DCO merger properties expected from different channels appears crucial for the science goals proposed in this section.
Currently, the metallicity dependence of the formation of DCO mergers (underlying some of those goals) is subject to uncertainties in stellar evolution models and lacks direct observational constraints - especially at low metallicity.
In turn, a great wealth of observational information about the properties of star forming galaxies already provides vital clues about the f$_{\rm SFR}$(Z,z) (see Section \ref{sec: fSFR(Z,z)}). 
Obtaining f$_{\rm SFR}$(Z,z) with well understood uncertainties in combination with improving constraints on the population properties of GW sources would yield valuable constraints on the evolutionary models and DCO merger formation scenarios.

\section{Metallicity dependent cosmic star formation history: where we stand}\label{sec: fSFR(Z,z)}

Given the apparent sensitivity of DCO (especially BH+BH) merger properties to metallicity, ignoring the uncertainty associated with f$_{\rm SFR}$(Z,z) in the calculations of the properties of the population of GW sources and in the subsequent comparison with GW observations may lead to erroneous conclusions.
It is thus crucial to establish the reasonable extremes of f$_{\rm SFR}$(Z,z), especially of its low metallicity part.
The related challenges and current efforts are summarized in Section \ref{subsec: f_SFR(Z,z) uncertainty}.
Section \ref{subsec: lit assumptions} offers a brief comparison of different f$_{\rm SFR}$(Z,z) used in the literature.

  \begin{figure}[h!]
  \begin{minipage}[c]{0.7\textwidth}
    \includegraphics[width=1.\textwidth]{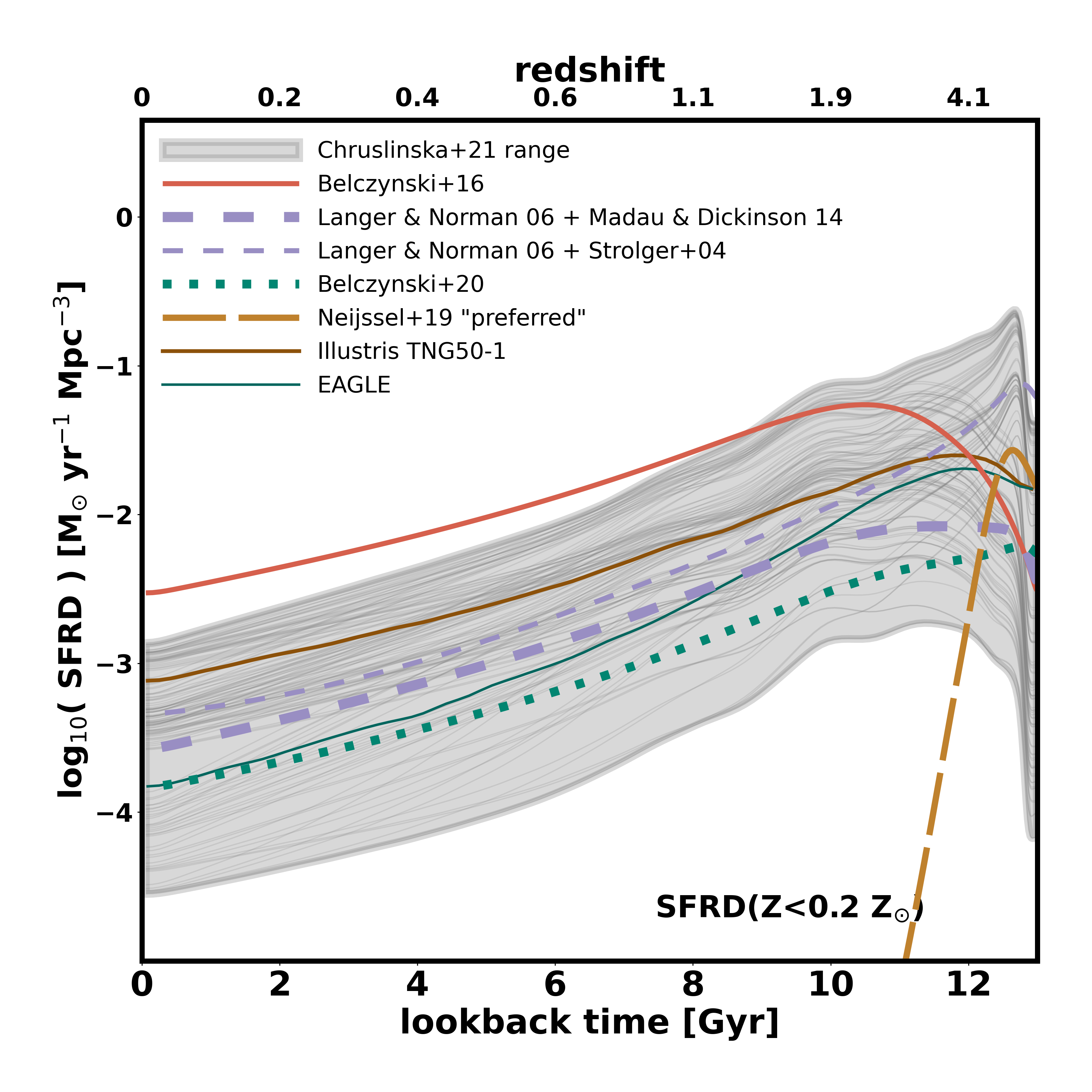}
  \end{minipage}\hfill
  \begin{minipage}[c]{0.3\textwidth}
    \caption{
    Low metallicity part of the cosmic SFH (SFRD happening below 1/5 solar metallicity at different times/redshifts) 
    based on a variety of methods:  cosmological simulations of galaxy evolution (EAGLE \cite{McAlpine16} and Illustris-TNG \cite{Nelson19_TNG, Pakmor22}), analytical prescriptions  (\cite{LangerNorman06} with birth metallicity distribution of stars normalized to the total SFRD as found by either \cite{MadauDickinson14} or \cite{Strolger04}; \cite{Belczynski16,Belczynski20}), fit of the population synthesis model to GW-based constraints \cite{Neijssel19} and the observation-based estimates \cite{Chruslinska21}. The gray range spans between the estimates from \cite{Chruslinska21}, while the thin gray lines show particular variations of their observation-based model (those estimates use $\rm Z_{O/H}$ based metallicity).
    See Section \ref{subsec: lit assumptions}.
    }             
    \label{fig: lowZ_compared}
  \end{minipage}
\end{figure}

\subsection{Understanding the uncertainty}\label{subsec: f_SFR(Z,z) uncertainty}

 There is no single, simple measurement providing constraints on the number and metallicity of stars forming across the entire Universe - f$_{\rm SFR}$(Z,z) needs to be inferred indirectly. 
This requires combining many pieces of information, all of which come with associated uncertainties and biases (see Section 2 in \cite{ChruslinskaNelemans19}).
Careful determination of the overall uncertainty is a challenge in itself.
\\
To estimate the f$_{\rm SFR}$(Z,z), it is necessary to know
the time evolution of the composition of matter that remains within galaxies and becomes available as the star forming material - 
i.e.\ one needs to understand the chemical evolution of galaxies.
Individual galaxies show a great variety of chemical and star formation histories,
depending, for instance, on their mass, environment in which they reside
and merger history.
However, when a representative population is considered,
the  average properties of star forming galaxies (their masses, star formation rates and metallicities) appear to be linked through relatively tight and simple relations \cite{Tremonti04,Brinchmann04,Ellison08}.
Those empirical relations can be used to obtain a observation-based estimate of  f$_{\rm SFR}$(Z,z)
without the need for a detailed description of the evolution of individual galaxies \cite{ChruslinskaNelemans19,Boco21,Chruslinska21}.
However, differences in the methods used to estimate the galaxy properties lead to a great variety of empirically derived relations.
This is particularly striking in case of metallicity derivations \cite{KewleyEllison08,MaiolinoMannucci19,Cresci19}.
It is often not possible to determine which (if any) of the applied techniques leads to the correct estimate.
Furthermore, observational results are increasingly uncertain/incomplete 
with increasing redshift and decreasing galaxy luminosity, 
in which regimes one has to rely on extrapolations.
Both factors introduce considerable uncertainty in the observation-based f$_{\rm SFR}$(Z,z) derivation \cite{Chruslinska21}.
Note that the star formation at low metallicity is thought to happen primarily in young (high-redshift) and low mass (faint) galaxies - i.e.\ exactly in those objects that are the most difficult to observe and whose properties are the most uncertain. 
Not surprisingly, there is a wide range of observationally-allowed estimates of the SFRD happening at low metallicity even at low redshifts (see \cite{Chruslinska21} and the gray band in Figure \ref{fig: lowZ_compared}). 
Finally, some factors that are typically not accounted for in the SFR derivation (for instance, the presence of interacting binaries \cite{Wilkins19}, stellar initial mass function variations \cite{Chruslinska20} - both affecting the interpretation of galaxy spectra, e.g. \cite{EldridgeStanway22}) introduce additional (systematic) uncertainty to f$_{\rm SFR}$(Z,z) that needs to be better quantified in the future.
\\
Alternative method to construct f$_{\rm SFR}$(Z,z) relies on the results of cosmological simulations/models of galaxy formation and evolution.
Cosmological simulations follow the formation of structures in a representative volume of a virtual universe calibrated to satisfy various observational constraints \cite{Vogelsberger14,Schaye15,McAlpine16,Nelson19_TNG,Pillepich18}.
As such, they provide a particular realization of the galaxy population that may exist in the real Universe.
% , but themselves do not allow to learn about the uncertainty of the resulting f$_{\rm SFR}$(Z,z)
Note that the differences between the results of current simulations do not cover the variety of observationally allowed f$_{\rm SFR}$(Z,z) \cite{Pakmor22}.
However, it is important to realize that in principle they have the flexibility to do so by varying parameters of simplified prescriptions used to describe processes that are not resolved (e.g.\ star formation, metal enrichment due to stellar winds and different types of supernovae, energy and momentum injected into the interstellar medium through outflows and radiation related to the supernovae and AGN activity \cite{Crain15} - all of those shape the chemical evolution of individual galaxies).
 Ultimately, both simulations and observations should provide a consistent answer.

\subsubsection{Major issues}\label{subsec: major issues}
There are many factors that limit our ability to accurately (either theoretically or observationally) determine the f$_{\rm SFR}$(Z,z) \cite{ChruslinskaNelemans19,Chruslinska20,Chruslinska21} and it is impossible to review all of them here. Instead, I present a (necessarily biased) selection of the related major open problems:
\begin{itemize}
    \item What is the metallicity evolution (especially at high redshift)?\\
    The metallicity of star forming material can be inferred primarily from (i) spectra of massive stars, (ii) the part of the galaxy spectrum that is affected by massive star emission or (iii) emission of HII regions ionised by massive stars (see \cite{MaiolinoMannucci19,Kewley19} for recent reviews of the methods). 
    Option (i) is only accessible in the nearby Universe. 
    Option (ii) is mostly limited to redshift$\sim$3 galaxies for which the (redshifted) rest-frame UV part of the spectrum can be studied with optical instruments. The high spectral resolution and signal required for a meaningful measurement makes it challenging to apply this technique on representative samples of objects (but see \cite{Strom22}).
    Option (iii), which is the most common technique, relies on rest-frame optical emission lines, which in turn are no longer accessible to optical instruments for galaxies with redshifts above z$\sim$3 (but note that they will be accessible to JWST \cite{Rieke19}).
    This means that at present, star-forming metallicities are essentially unconstrained at redshifts $>$3.\\
    Furthermore, only metallicity estimates obtained with (iii) are currently available for large galaxy samples across redshifts.
    However, recent studies show that important adjustments to applied metallicity calibrations are necessary when galaxies at different redshifts are considered, questioning the overall metallicity evolution resulting from earlier studies (which likely overestimate the rate of that evolution - milder evolution is also expected from theoretical models/simulations).\\
    This issue is one of the biggest sources of uncertainty for f$_{\rm SFR}$(Z,z) (see e.g. \cite{Boco21,Chruslinska21}).
    \item How to constrain the abundance of iron in the star forming material?\\
The evolution of massive stars (and of the related GW source progenitors) is particularly sensitive to the abundance of one particular element - iron (see Section \ref{subsec: metallicity and stellar evolution}).
However, the most common observational metallicity estimates characterizing the star-forming material (i.e.\ option (iii) mentioned above) rely on the measurements of the oxygen abundance ($\rm Z_{O/H}$). 
Those two elements are released to the interstellar medium of galaxies by different sources and evolve on different timescales: oxygen is primarily released by core-collapse supernovae within a few 10 Myr after the star formation, while iron is abundantly released by type Ia supernovae with significant delay ($\sim$1 Gyr) with respect to star formation \cite{Tinsley79,Nomoto13}.
Therefore, oxygen-based metallicity is a very bad proxy for iron abundance already at redshift$<$2.
It is not straightforward to translate one metallicity measure to the other. 
Currently, the two are simultaneously estimated (for star forming material) only for a limited sample of galaxies at high redshift \cite{Steidel16,Cullen21,Strom22}.
Those galaxies reveal a significantly higher oxygen enhancement relative to iron than expected in the nearby Universe.
Consequently, current observation-based f$_{\rm SFR}$(Z,z) models (that to large extent rely on oxygen-based metallicity estimates) only provide a lower limit on the amount of the low (iron-based) metallicity star formation at high redshift.
This is a important caveat that should be kept in mind when those results are applied to model the populations of GW sources (and other transients related to massive stars).
\item What is the contribution of low mass galaxies to the total SFRD (and to low metallicity star formation)?
\\
The properties of low mass galaxies (with stellar mass $<$10$^{8}$ M$_{\odot}$) are poorly constrained, as such objects typically fall below the sensitivity/completeness limits of galaxy surveys (especially at higher redshifts; see Section 2 in \cite{ChruslinskaNelemans19} and references therein).
Note that also cosmological simulations typically lack resolution to robustly describe the population of such low mass galaxies.
Galaxy luminosity/mass function determinations hint at the potential strong redshift evolution of the faint/low mass slope (suggesting that the number and/or star formation rate of such galaxies increases with redshift). If such a trend is correct, those galaxies can dominate the total star formation rate budget and determine the shape of f$_{\rm SFR}$(Z,z) at redshifts$>$4, shifting the peak of the birth metallicity distribution of stars to very low values \cite{ChruslinskaNelemans19,Chruslinska21}. 
Poorly constrained properties of the low mass galaxies are also one of the main sources of uncertainty of the f$_{\rm SFR}$(Z,z) (especially its low metallicity part) in the local Universe.
Constraining their contribution is thus particularly important in the context of GW astrophysics. 
\item What is the contribution of starbursts (galaxies with abnormally high star formation) to the low metallicity star formation?
\\
Recent studies report a significantly higher fraction of starburst galaxies among the star forming population than previously found \cite{Caputi17,Bisigello18,Rinaldi21}.
Furthermore, their prevalence seems to increase towards low galaxy masses and high redshifts.
There is limited direct information about the metallicity of those galaxies. However, 
given that the star formation rate and metallicity of star forming galaxies appears to be anti-correlated, one may expect that starbursts (significantly) contribute to star formation at relatively low metallicity (see \cite{Chruslinska21} and references therein).
Note that galaxies with such properties are lacking in the current cosmological simulations \cite{Sparre15,Katsianis21,Rinaldi21}.
This should be kept in mind when their results are applied to model f$_{\rm SFR}$(Z,z) and used to characterize the populations of GW sources.
\end{itemize}

\subsubsection{Notes on a  realistic f$_{\rm SFR}$(Z,z) model}\label{subsec: realistic model}

While the exact characteristics of the f$_{\rm SFR}$(Z,z) are uncertain, both observation-based and cosmological simulation-based approaches lead to f$_{\rm SFR}$(Z,z) estimates that show some general features that any realistic model should reproduce (see Section 4 in \cite{ChruslinskaNelemans19} for more details).
Some of the simplified analytical f$_{\rm SFR}$(Z,z) prescriptions used in the literature fail to match those characteristics.
Most importantly, birth metallicity distribution of stars at any given redshift is not symmetric, but shows an extended low metallicity tail\footnote{
To first order, the existence of this asymmetry can be understood in light of the typical properties of star forming galaxies: for any given mass and redshift galaxies that are younger (i.e. their interstellar is still relatively metal-poor) or rejuvenated by (typically more metal-poor) gas accretion that can fuel further star formation typically have higher SFR (which is observationally manifested as the "fundamental metallicity relation", also expected in the models/cosmological simulations e.g. \cite{Ellison08,Mannucci10,Yates12,Torrey18}). The asymmetry is further strengthened by the (observed) flattening of the mass-metallicity relation at high masses (metallicities) and at high redshifts - by the increasing contribution of low mass galaxies to the total SFRD budget. See Section 3 in \cite{Chruslinska21} for further discussion and references.
} 
(even at low redshifts, there is some level of very low metallicity star formation - relevant for BH+BH/BH+NS mergers; e.g. \cite{ChruslinskaNelemans19, Boco21, Chruslinska20, Chruslinska21,Briel21, Pakmor22} ).
Note that the empirical metallicity measures (in particular $\rm Z_{O/H}$ used as a metallicity proxy in the observationa-based f$_{\rm SFR}$(Z,z) estimates) use logarithmic scales.
In this view, the parametrisation proposed by \cite{LangerNorman06} is a more appropriate choice than a log-normal distribution (used e.g.\ in \cite{Belczynski16,Belczynski20,Santoliquido20}, and the preferred model from \cite{Neijssel19}).
Furthermore, such simplified prescriptions are often constructed assuming that the total SFRD (f$_{\rm SFR}$(Z,z) integrated over metallicity at each redshift) and the metallicity distribution are independent, and often consider variations combining the same birth metallicity distribution of stars with different total SFRD determinations (e.g.\ \cite{Neijssel19,Tang20,Belczynski20,Broekgaarden22}).
However, not all such variations are realistic.
Uncertainties in the total SFRD stem primarily from difficulties with constraining the contribution of particular types of galaxies (e.g.\ massive, highly star forming galaxies that are severely obscured by dust or faint, low mass objects).
Given the fact that the properties of galaxies are correlated, those objects contribute to star formation only in a particular range of metallicities.

\subsection{Variety of assumptions}\label{subsec: lit assumptions}
Various methods (based on observations, simulations, analytic prescriptions or on some combination of those approaches) were used to establish f$_{\rm SFR}$(Z,z) and applied in the literature to estimate the properties of GW sources.
Figure \ref{fig: lowZ_compared}
shows an incomplete comparison of those assumptions.
Specifically, it shows the amount of star formation happening at low metallicity (below 1/5 solar metallicity - the chosen metallicity threshold is arbitrary) as a function of time/redshift.
Note that the presented f$_{\rm SFR}$(Z,z) also significantly differ in shape and in the total SFRD as a function of redshift - such differences are not captured in the simple comparison presented in Figure \ref{fig: lowZ_compared} (in particular, the fact that the predicted amount of star formation happening at low metallicity is reasonable does not imply that the overall f$_{\rm SFR}$(Z,z) is correct - see Section \ref{subsec: realistic model}).
Overall, those estimates differ by orders of magnitude (at any redshift), have a variety of slopes and peak locations.
In-depth comparison of the different assumptions is beyond the scope of this paper (but see Section 6.3 in \cite{ChruslinskaNelemans19} for a detailed discussion of some of those models).
The most extreme assumption shown in Figure \ref{fig: lowZ_compared}, clearly standing out from all the other estimates, is the 'preferred' model from \cite{Neijssel19}. 
Interestingly, it assumes an analytical form for the f$_{\rm SFR}$(Z,z) with parameters chosen in a way that allows to satisfy GW-based constraints (reported after the first two LIGO/Virgo observing runs \cite{GWTC2pop}) with their fiducial population synthesis model (not surprisingly, this f$_{\rm SFR}$(Z,z) allows to match the GW-based constraints shown in Figure \ref{fig: rate example}). The fact that this f$_{\rm SFR}$(Z,z) model predicts negligible low metallicity SFR below redshift $<$2, in tension with observation-based estimates (which likely underestimate this quantity - see Section \ref{subsec: major issues}), might already signal that the adopted evolutionary assumptions overestimate the formation efficiency of BH+BH mergers (see also \cite{Gallegos-Garcia21}).

\section{Interpretation challenge and the role of future detectors} \label{subsec: degeneracies and 3G}

A great variety of assumptions about both the evolution leading to the formation of DCO mergers and f$_{\rm SFR}$(Z,z) is currently used in the literature.
The example considered in Section \ref{subsec: BHBH mergers} demonstrates that if the low metallicity part of the cosmic SFH is very uncertain, this by itself adds significant uncertainty to the population properties of BH+BH mergers.
In particular, if both of the compared f$_{\rm SFR}$(Z,z) are realistic, one cannot judge the correctness of the assumed evolutionary model. 
Furthermore, it shows that unrealistic f$_{\rm SFR}$(Z,z) model can easily lead to erroneous conclusions.
This poses a serious problem for the interpretation of GW observations, further complicated by the unknown contribution of DCO mergers formed in non-isolated channels.
In order to alleviate those issues, future studies should aim to:\\
(i) explore ways to obtain tighter constraints on the key components of the population modelling, especially in the low metallicity regime.
It is important to maintain the dialogue between the fields of GW astrophysics and galaxy chemical evolution to make further progress in constraining f$_{\rm SFR}$(Z,z).
\\
(ii) take f$_{\rm SFR}$(Z,z) uncertainty into account during the comparison with observations. Relying on one particular f$_{\rm SFR}$(Z,z) model can drastically bias the results. Considering a number of f$_{\rm SFR}$(Z,z) spanning the observationally-allowed range in combination with evolutionary models would allow for a more fair comparison with the GW-based constraints.
Furthermore, releasing the evolutionary models (before combining them with f$_{\rm SFR}$(Z,z)) would allow one to bring the results to a common baseline and compare or combine with the updated f$_{\rm SFR}$(Z,z).\\
(iii) identify ways to break degeneracies between the f$_{\rm SFR}$(Z,z) and metallicity-dependent stellar evolution related assumptions in the GW-observable properties of the population of DCO mergers.
\\
One way to move forward with (iii) is to consider all GW-based constraints (merger rate density, mass/mass ratio/spin distributions, stochastic GW background from unresolved mergers) simultaneously when comparing the models with observations. 
Stochastic GW background is rarely discussed in this context, with notable exceptions of \cite{Callister20,Bavera22}.
For current GW detectors, this signal is dominated by DCO, in particular BH+BH mergers \cite{Abbott21_SGWB}. 
It provides a limit on the net rate of those mergers across all redshifts, allowing to peek into the early cosmic history before the era of third generation GW detectors.
When combined with information from direct detections (currently limited to redshifts $<$1), it provides stronger constraints on the BH+BH merger rate as a function of redshift \cite{Callister20,Abbott21_SGWB}.\\
Spin distribution is currently mostly discussed as a potential method to distinguish between the formation channels (e.g. \cite{GWTC3}).
However, in case of isolated binary evolution it may also depend on metallicity and delay time \cite{Bavera22spin} and therefore probe f$_{\rm SFR}$(Z,z).
  \begin{figure}[h!]
  \begin{minipage}[c]{0.5\textwidth}
    \includegraphics[width=1.\textwidth]{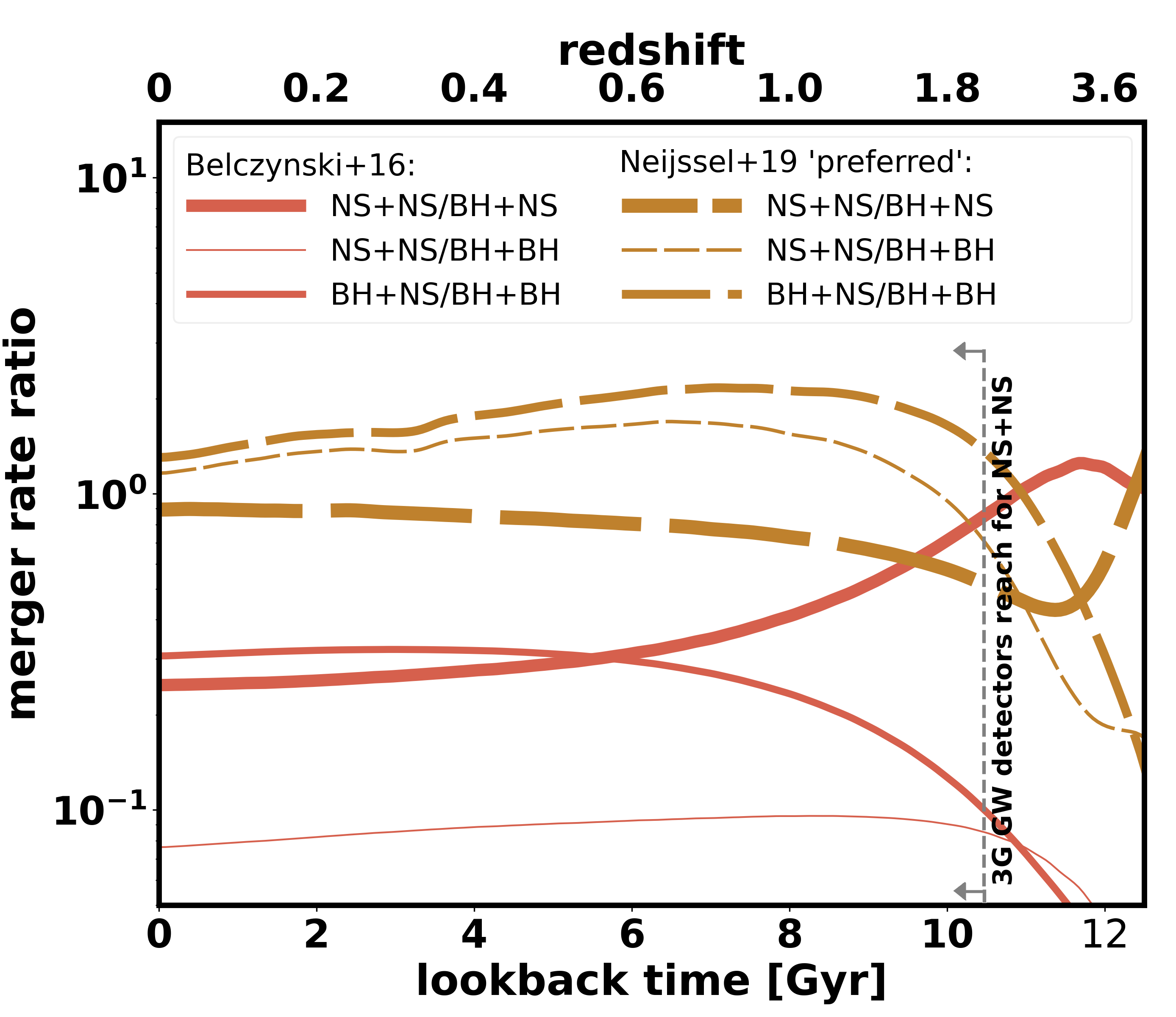}
  \end{minipage}\hfill
  \begin{minipage}[c]{0.5\textwidth}
    \caption{Ratio of the merger rate of different DCO (see legend) as a function of lookback time/redshift for the same population synthesis model and two (very) different models of metallicity-dependent cosmic SFH as used in Figure \ref{fig: rate example}.
    The vertical dashed line indicates the expected reach of the proposed third generation GW detectors for NS+NS mergers \cite{BorhanianSathyaprakash22} (see text; more massive BH+NS and BH+BH systems will be seen to further redshifts).
    }             
    \label{fig: rate ratio example}
  \end{minipage}
\end{figure}
While the current GW-based limits on spins and stochastic GW background are not particularly constraining, they will continue to improve and it is important to prepare the models for their interpretation.
\\
With the improving sensitivity of current and future detectors it will become possible to observe DCO mergers happening at higher redshifts \cite{Punturo10,Reitze19,Maggiore20}.
\cite{BorhanianSathyaprakash22} estimate the reach (here defined as redshift $z_{r}$ up to which at least half of the mergers of a given type would be detected with signal to noise ratio $>$10) of the combined network of advanced detectors that could be operational in the next 5-10 years (see Table 3 therein) to be
$z_{r}$=0.11 and $z_{r}$=0.6 for NS+NS and BH+BH mergers respectively. In turn, the most optimistic configuration of three next generation GW detectors considered in their study yields $z_{r}\sim$2 for NS+NS and $z_{r}>$40 for BH+BH mergers. This would allow to fully map the redshift evolution of BH+BH merger population properties and probe the properties of the NS+NS mergers in the bulk of the cosmic SFH.
Some diagnostics that take advantage of this fact
to disentangle the effects of f$_{\rm SFR}$(Z,z) and evolution/DCO formation scenarios
% in this context 
are discussed below.
While individually each of them is affected by both factors, when considered jointly they can help to break the degeneracies.
\\
(i)  Individual merger rates as a function of redshift:
rate evolution may be compared with the total cosmic SFH (i.e. f$_{\rm SFR}$(Z,z) integrated over all metallicities, which is better constrained than the f$_{\rm SFR}$(Z,z))
to gain insights about the joint effect of metallicity dependence of the DCO formation and merger delay time distribution, e.g.\cite{Fishbach21}. 
BH+BH merger rate peak location alone may help to discriminate between extreme f$_{\rm SFR}$(Z,z) models (see Figure \ref{fig: rate example}).
\\
(ii) Redshift evolution of the merger rate ratios of different DCO mergers: the fact that the formation of NS+NS, NS+BH and BH+BH mergers likely shows different metallicity dependence makes the redshift evolution of the ratio of their merger rates particularly sensitive to f$_{\rm SFR}$(Z,z). This is illustrated in Figure \ref{fig: rate ratio example} for the same example models as considered in Section \ref{subsec: BHBH mergers}.
While the redshift evolution of BH+BH mergers may be severely impacted by the contribution of different formation channels, this is less likely for mergers containing NS (e.g.\ \cite{Belczynski18,Zevin19,Ye20}).
 The evolution of the relative rate of NS+NS and NS+BH mergers is thus a promising tool that can help to disentangle the effects of stellar evolution and f$_{\rm SFR}$(Z,z).
\\
(iii) features in the BH mass distribution:
given the fact that the mass distribution of compact objects formed in stellar evolution depends on metallicity, different features/peaks in the BH mass distribution evolve with redshift in a way that depends on f$_{\rm SFR}$(Z,z).\\
Particular example is the predicted pileup in the BH mass distribution related to (pulsational) pair-instability supernovae (e.g. \cite{Stevenson19,vanSon22}).
Its location is relatively insensitive to metallicity \cite{Farmer19}. At the same time, only BH formed from sufficiently metal poor progenitors can contribute to it.
Once identified in the GW-observed mass distribution, the redshift evolution of this feature is thus a promising probe of f$_{\rm SFR}$(Z,z).
Similarly, one could use any potential feature in the spin distribution that can be related to metallicity as a 
f$_{\rm SFR}$(Z,z) diagnostic.
Conversely, features that are not sensitive to metallicity will only depend on the evolution/DCO formation and the total cosmic SFH.\\
Certain model assumptions (either about f$_{\rm SFR}$(Z,z) or DCO formation) may strongly affect the properties of the DCO merger population only at high redshifts $>2$ and beyond.
It is important to address the question to what extent the uncertainties associated with the future GW observations (due to sensitivity limitations and the necessity to deal with overlapping GW events, e.g. \cite{Relton21}) at those redshifts will allow to discriminate such models (see e.g. \cite{Singh21,Pieroni22,Yi22} for the recent discussion).

\section{Conclusions}

The metallicity dependent cosmic star formation history is an equally important ingredient of the modelling of the populations of GW sources as the description of the evolution leading to the formation of merging systems:
\begin{itemize}
\item 
 Depending on the assumed f$_{\rm SFR}$(Z,z) one can obtain significantly different properties (merger rate, relative rates of different types of DCO mergers, masses) of the model population as a function of redshift.
 
\item 
 Different combinations of assumptions about the evolution leading to the formation of DCO mergers and about the f$_{\rm SFR}$(Z,z) can lead to very similar properties of the model population of mergers in the redshift range probed by current GW detectors, hindering the meaningful interpretation of those observations.
 
 \item 
 Especially the low metallicity part of f$_{\rm SFR}$(Z,z) is currently poorly constrained even at low redshifts (see Sections \ref{subsec: f_SFR(Z,z) uncertainty} and \ref{subsec: major issues}).
 This by itself can add significant uncertainty to the modelled properties of DCO mergers involving BH (also in the models predicting short merger times),
  whose formation efficiency strongly favors low metallicity according to current models (see Section \ref{subsec: BHBH mergers}).
% even in the models predicting that the progenitors of current mergers formed within the last 1 Gyr.
\end{itemize} 
 
Even though the importance of the f$_{\rm SFR}$(Z,z) assumption is now well established in the literature, the associated uncertainty is rarely discussed.
Ignoring this uncertainty in the comparison of the model predictions with observations can lead to erroneous or overstated conclusions.\\
One of the biggest challenges during the interpretation of the observed properties of the populations of GW sources is the presence of degeneracies in the model predictions.
Future generations of GW detectors will constrain the properties of GW sources as a function of redshift. 
This will provide additional information which can be used to disentangle the effects of evolution/different formation scenarios and f$_{\rm SFR}$(Z,z) on the properties of DCO mergers and break such degeneracies (see Section \ref{subsec: degeneracies and 3G}). To prepare for this opportunity, future studies should aim to:
\begin{itemize}
    \item explore the redshift dependent DCO merger rate, mass (ratio) and spin distributions for different formation channels and consider a range of metallicity-dependent star/cluster formation histories allowing to discuss the uncertainties related to this part of the modelling.
    \item consider all GW-observable quantities related to DCO merger population simultaneously.
\end{itemize}

However, if both the f$_{\rm SFR}$(Z,z) and the evolutionary part of the modelling remain poorly constrained, this may still challenge the meaningful interpretation of even those future GW observations.
Studies modelling the formation of DCO mergers should aim to:
\begin{itemize}
    \item Better quantify model uncertainties and, where possible, establish realistic ranges of assumptions about the major parameters. 
    \item Indicate which part(s) of the parameter space cannot be explained with a given model 
    \item Verify and update the simplified prescriptions implemented in rapid codes (used to model DCO merger populations) with detailed evolutionary models across metallicities.
\end{itemize}
Given the huge parameter space involved and the fact that such models often combine input from various sub-fields of astrophysics (with all their related uncertainties and limitations), the first is easier said than done.
 In this view, quantifying what a given model cannot predict seems equally (if not more) important as demonstrating that a particular combination of assumptions can match the observed population properties.
 The last point is crucial to better understand the robustness of the currently found metallicity (in)sensitivity of the efficiency of formation of (NS+NS)  BH+BH mergers.
    Recent detailed calculations already show that some relevant metallicity-dependent effects may be missing in the rapid codes (see Section \ref{subsec: evolution effects on Xeff}). 
 \\
Finally, a few points related to the modelling of the metallicity-dependent cosmic star formation history are worth emphasizing:
\begin{itemize}
    \item Despite the difficulties with constraining the f$_{\rm SFR}$(Z,z), at least some of its general properties are already well understood (see Section \ref{subsec: f_SFR(Z,z) uncertainty}) - any reasonable f$_{\rm SFR}$(Z,z) prescription should aim to reproduce those characteristics.
    \item Current and future facilities and surveys targeting galaxies at redshifts $>$2 and aiming to considerably increase the sample of galaxies with metallicity measurements (e.g.\ \cite{MOONRISE,Rieke19}) will allow to improve the overall constraints on this quantity.
    \item As long as metallicity estimates for representative sample of galaxies are limited to oxygen abundance tracers, a source of f$_{\rm SFR}$(Z,z) uncertainty that is particularly important for studies modelling the populations of objects related to massive stars - i.e.\ uncertain iron abundance - will remain.
Constraints on the multi-element abundances (including both oxygen and iron) in star forming material are currently limited. Expanding in this direction appears crucial in the context of GW astrophysics.
    \item  Galaxies shaping the low metallicity tail of f$_{\rm SFR}$(Z,z) (likely to be faint and high redshift) will always pose a challenge to electromagnetic observations. 
 The apparent sensitivity of BH+BH merger properties to low metallicity SFH offers an exciting possibility of constraining the properties of such galaxies with future GW observations.
\end{itemize}
It is important to note that currently there is very limited direct observational information that can be used to constrain the models of evolution of massive stars at metallicities lower than those of the nearby Small and Large Magellanic Clouds. 
At the same time, various energetic transients linked to massive star/binary progenitors (e.g. long gamma ray bursts \cite{Kruhler15,Perley16}, hydrogen-deficient superluminous supernovae \cite{Moriya18}, ultraluminous X-ray sources \cite{Kovlakas20}) are preferentially found in metal-poor environments.
While the strong low metallicity preference found theoretically for BH+BH mergers should be verified in future studies (incorporating results of detailed models), it is not an isolated clue of the importance of metal-poor regime for massive star evolution.

\medskip
% \textbf{Supporting Information} \par %Please delete the Suppporting Information statement if it is not applicable. Please supply Supporting Information in another file. Supporting information should not be provided in .tex format
% Supporting Information is available from the Wiley Online Library or from the author.

% Acknowledgements
\medskip
\textbf{Acknowledgements} \par %delete if not applicable))
I wish to thank Selma de Mink, Jakub Klencki and Stephen Justham for their comments.
I am grateful to anonymous referees for their comments and suggestions.
% References
\medskip

% Use the following code if you wish to generate your bibliography with BibTeX;
% replace the string "MSP-template" below with the name(s) of
% the BibTeX data base(s) you want to use.
% The resulting bibliography-output (the content of the .bbl file)
% must be pasted back into this file before submission.
% Please also include your BibTeX data base file(s) in your submission
% so that we can re-run BibTeX if necessary.
%
 \bibliographystyle{MSP_modif}
 \bibliography{main}

\end{document}